\input{aipcheck}

\documentclass[
  ,final            
  ]
  {aipproc}

\layoutstyle{8x11single}

\begin{document}

\title{Temporal variability of GRB early X-ray afterglows and GRB080319B prompt emission}


\author{R. Margutti}{
address={INAF, Osservatorio Astronomico di Brera, Via E. Bianchi 46, I-23807, Merate (LC), Italy }
  ,altaddress={Universit\'a degli Studi di Milano-Bicocca, Dipartimento di Fisica, Piazza della Scienza 3, I-20126 Milano, Italy. }
}

\author{C. Guidorzi}{
  address={INAF, Osservatorio Astronomico di Brera, Via E. Bianchi 46, I-23807, Merate (LC), Italy }
}

\author{G. Chincarini}{
  address={INAF, Osservatorio Astronomico di Brera, Via E. Bianchi 46, I-23807, Merate (LC), Italy }
  ,altaddress={Universit\'a degli Studi di Milano-Bicocca, Dipartimento di Fisica, Piazza della Scienza 3, I-20126 Milano, Italy. }
}

\author{F. Pasotti}{
address={INAF, Osservatorio Astronomico di Brera, Via E. Bianchi 46, I-23807, Merate (LC), Italy }
 ,altaddress={Universit\'a degli Studi di Milano-Bicocca, Dipartimento di Fisica, Piazza della Scienza 3, I-20126 Milano, Italy. }
}

\author{S.Covino}{
address={INAF, Osservatorio Astronomico di Brera, Via E. Bianchi 46, I-23807, Merate (LC), Italy }
}

\author{J. Mao}{
address={INAF, Osservatorio Astronomico di Brera, Via E. Bianchi 46, I-23807, Merate (LC), Italy }
 }

\begin{abstract}
We performed the first systematic search for the minimum variability time scale 
between
$0.3$ and $10$ keV studying the $28$ brightest early ($<3000$ s) afterglows
detected by Swift-XRT up to March 2008.
We adopt the power spectrum analysis in the time domain: 
unlike the Fourier spectrum, this is suitable to study
the rms variations at different time-scales.
We find that early XRT afterglows show
variability in excess of the Poissonian noise level on time-scales as short
as  $\sim 1$ s (rest frame value), with the shortest  $t_{min}$ associated with 
the highest energy band.

The $\gamma-$ray prompt emission of GRB\,080319B shows a characteristic average 
variability time-scale $<t_{var}> \sim 1 $s; this parameter undergoes
a remarkable evolution during the prompt emission (BAT observation). 
\end{abstract}

\maketitle

\section{introduction}
The time variability in afterglow and prompt emission
light curves can provide important
clues to the nature of the source that powers the Gamma Ray Burst (GRB) 
emission and, possibly, of its surroundings.

In this work we characterize the temporal variability properties 
of the detected GRB emission through 
its power density spectrum (PDS) in the time domain (see 
\cite{li:2001} and \cite{li:2002}). In particular, the variation
power $P(\Delta t)$ in a count light curve $x(t)$ as function of the 
binning time $\Delta t$ is defined as:
\begin{equation}
P(\Delta t)=\frac{Var(x)}{(\Delta t)^2}\,\,\,\,\,\,\,\,\,\,\,\,\,\,\rm{rms^2}
\end{equation}
From this quantity it is possible to derive the power density  
$p(\Delta t)$ in the time domain defined as the rate of change
of $P(\Delta t)$ with respect to the time step $\Delta t$:
\begin{equation}
p(\Delta t)=\frac{P(\Delta t_{1})-P(\Delta t_{2})}{\Delta t_{2}-\Delta t_{1}}\,\,\,\,\,\,\,\,\,\,\,\,\,\,\rm{rms^2 s^{-1}}
\end{equation}
where $\Delta t_{2}>\Delta t_{1}$ and $\Delta t=(\Delta t_{1}+\Delta t_{2})/2$.

For a pure Poisson noise distribution, the previous equation reduces to:
\begin{equation}
p_{noise}(\Delta t)=\frac{r}{\Delta t_{1} \Delta t_{2}}\,\,\,\,\,\,\,\,\,\,\,\,\,\,\rm{rms^2 s^{-1}}
\end{equation}
where $r$ is the mean observed count rate. In this way it is possible to define
the fractional signal power density (fpd)- i.e. the total fractional power 
density removed of the statistical noise-  and the power density ratio
(pdr) as follows:
\begin{equation}
\label{Eq:fpd}
fpd(\Delta t)=\frac{p(\Delta t)-p_{noise}(\Delta t)}{r^2}\,\,\,\,\,\,\,\,\,\,\,\,\,\,\rm{(rms/mean)^2s^{-1}}
\end{equation}
\begin{equation}
\label{Eq:rpd}
pdr(\Delta t)=\frac{p(\Delta t)}{p_{noise}(\Delta t)}
\end{equation}

\section{Timing Analysis}
\subsection{Early X-Ray afterglows}

\begin{table}
\begin{tabular}{llllllll}
\hline
\tablehead{1}{l}{t}{GRB}\\  
\hline
080319B&070616&061121&060814&060526&060210&060105\\   
080310&070419B&061007&060729&060510B&060202&051117A\\
080212&070328&060904B&060614&060418&060124&050730\\
071031&070129&060904A&060607A&060218&060111A&050724\\
\hline
\end{tabular}
\caption{The 28 brightest early ($t_{max}<3000$ s) X-ray afterglows detected by 
Swift-XRT up to March 2008 whose light curves are not affected by data
gaps.}
\label{Tab:unica}
\end{table}

\begin{figure}
\label{Fig:tmin}
  \includegraphics[scale=0.9]{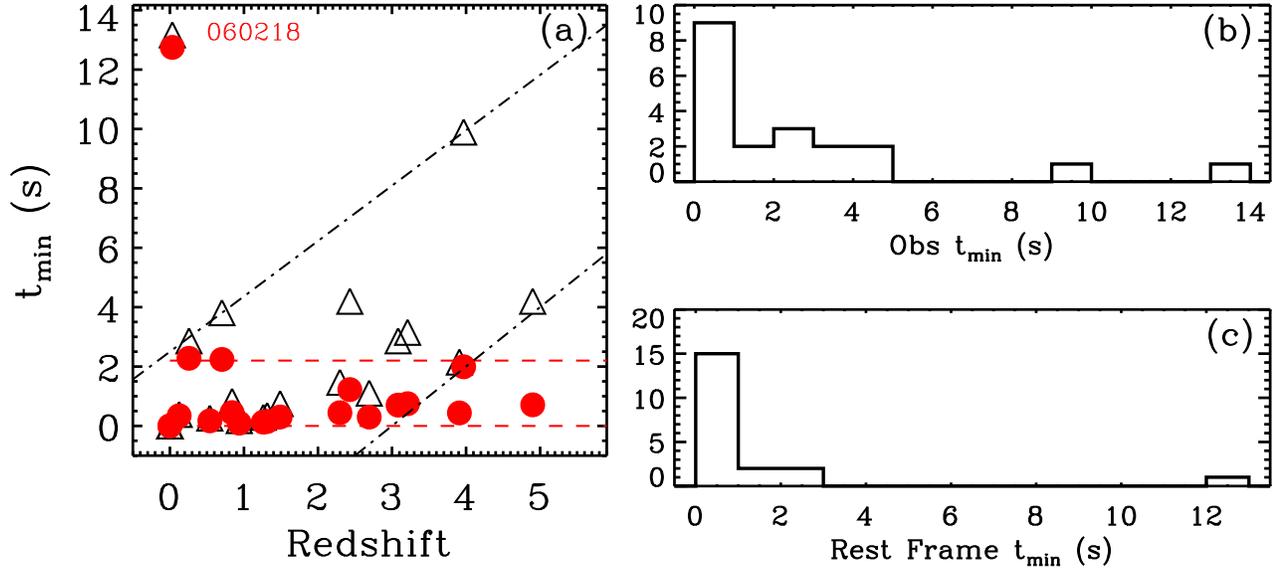}
  \caption{(a) Minimum detectable variability time scale $t_{min}$ as a function of redshift:
empty triangles: observed values; filled circles: redshift corrected values. No evolution of the $t_{min}$ parameter is apparent when the cosmological time dilation is properly considered (black dot-dashed lines vs. red dashed lines). Right panel:observed (b) and redshift corrected (c) $t_{min}$ distributions.}
\end{figure}

\begin{figure}
\label{Fig:tminEn}
  \includegraphics[scale=1.]{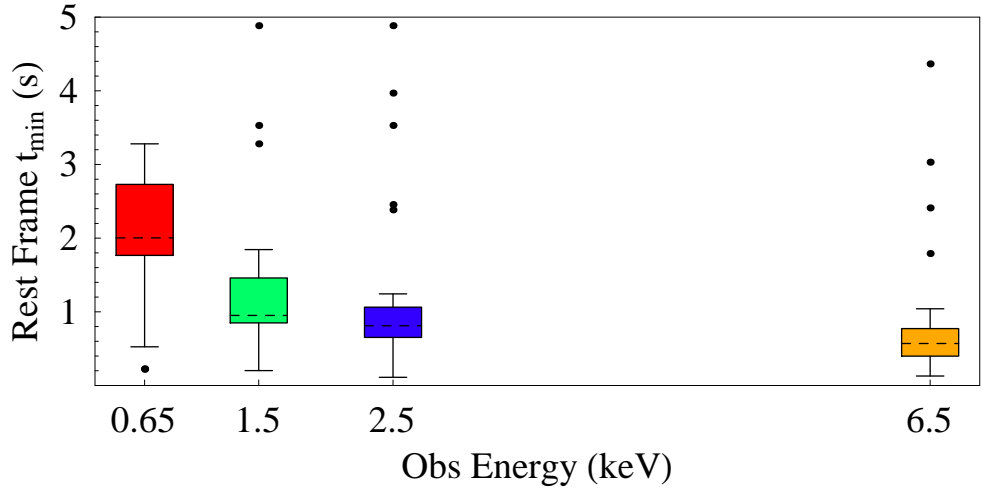}
  \caption{Box-and-whisker plot showing the rest frame minimum time scale of
variability as a function of the observed energy band of light curve extraction. 
Dashed lines: median values. Box edges:quantiles $40$\% and $60$\%. The thick black lines
mark the data set extension excluding outliers (filled black circles).}
\end{figure}

We studied the PDS in the time domain of the 28 
brightest early ($t_{max}<3000$ s) X-ray afterglows detected by 
Swift-XRT up to March 2008, with no observational gaps in their 
light curves (Table \ref{Tab:unica}). 
We refer the reader to 
Margutti et al. in prep. for details about the sample selection
and light curve extraction.
For each GRB $0.3-10$ keV  light curve, we calculated the pdr
(Eq.\ref{Eq:rpd}) and defined the minimum detectable variability
time-scale $t_{min}$ as the shortest time-scale showing a pdr above the
$3\sigma$ level expected from a pure Poisson noise distribution 
estimated via Monte Carlo simulations. Results are shown in Fig. \ref{Fig:tmin}:
most of early X-ray afterglows have $t_{min}<1$ s.
A definite exception is GRB\,060218 with $t_{min}>10$ s. 
While no evolution with redshift is detected (Fig.\ref{Fig:tmin}),
$t_{min}$ shows instead a clear trend with energy: the softer the
energy band, the longer the variability time scale
(see Fig. \ref{Fig:tminEn}). 
Observational effects would eventually work against this and therefore
can only strengthen this conclusion.

\begin{figure}
\label{Fig:080319B}
  \includegraphics[scale=0.7]{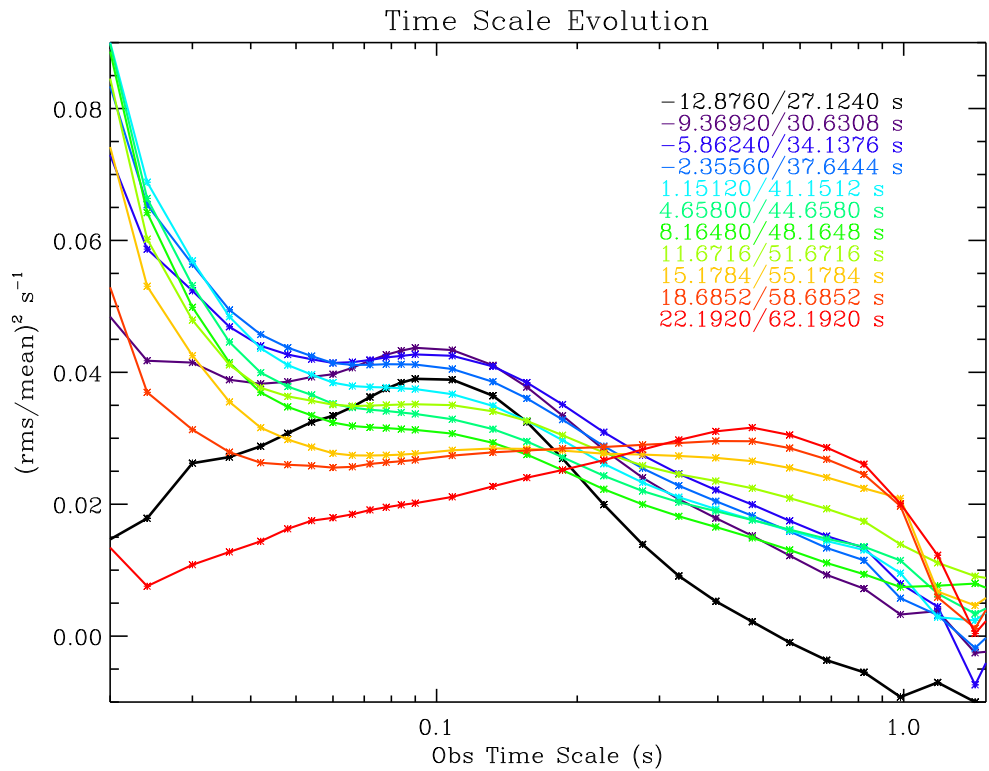}
  \caption{Time resolved analysis of GRB\,080319B prompt emission ($15-150$ keV). The
fpd defined as in Eq.\ref{Eq:fpd} is shown as a function of the observed time-scale: 
Color coding corresponds to a $40$ s long interval moving along
the GRB profile by steps of $3.5$s.
Observer frame time intervals are listed.}
\end{figure}

\subsection{GRB\,080319B prompt emission}
The PDS  of GRB\,080319B prompt emission ($15-150$ keV) has been 
calculated considering the Gaussian nature
of the noise distribution (see \cite{rizzuto:2007} for details).
In particular, we study in this case the fpd (Eq.\ref{Eq:fpd}):
this quantity is expected to show a peak whenever
a characteristic time-scale of variability is encountered (see e.g. 
\cite{shen:2003}).

\begin{figure}
\label{Fig:080319Bbis}
  \includegraphics[scale=0.6]{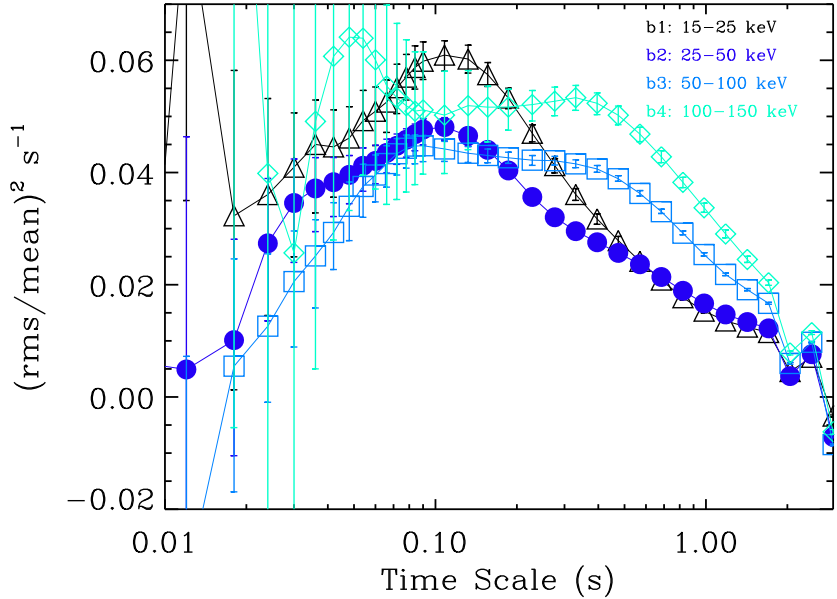}
  \caption{Energy resolved analysis of GRB\,080319B prompt emission. The
fpd defined as in Eq.\ref{Eq:fpd} is shown as a function of the observed time-scale.
Triangles: $15-25\,\rm{keV}$; circles: $25-50\,\rm{keV}$;
squares:$50-100\,\rm{keV}$; diamonds:$100-150\,\rm{keV}$.}

\end{figure}

While the total GRB\,080319B profile shows a variability time-scale
in the range $\sim0.1-1$ s, an interesting result is obtained if we do a time-resolved analysis
of the light curve. From Fig. \ref{Fig:080319B} is apparent
that $t_{var}$ undergoes a significant change of about one order of magnitude
from the beginning to the end of the detected emission. In particular,
while the first $40$s of the light curve are dominated by $t_{var}\approx0.1$s,
the last part shows a much longer characteristic time-scale $t_{var}\approx0.7$s. 
Moreover it is also apparent that there is not a continuous shift from
the $0.1$s to the $0.7$s time-scale. What we do observe is instead 
a progressive depletion of the fractional power associated with the shorter time-scale
and the contemporary rise of the $0.7$s fpd.
Finally, an energy resolved analysis of the same temporal profile 
(see fig. \ref{Fig:080319Bbis}) reveals
that the presence of two distinct time scales is peculiar of the 
highest energy band ($100-150\, \rm{keV}$), the lowest one ($15-25\, \rm{keV}$)
showing a unique variability time-scale $t_{var}\approx0.1$s.
We refer the reader to Guidorzi et al. in prep. for a complete discussion
of this topic and a possible physical interpretation.

\begin{theacknowledgments}
This work is supported  by ASI Grant ASI grant I/011/07/0, by the MIUR grant 2005025417) and by the University of Milano Bicocca (Italy).
\end{theacknowledgments}



\bibliographystyle{aipproc}   

\bibliography{Marguttiv0}

\hyphenation{Post-Script Sprin-ger}
\begin{thebibliography}{4}
\expandafter\ifx\csname natexlab\endcsname\relax\def\natexlab#1{#1}\fi
\providecommand{\enquote}[1]{``#1''}
\expandafter\ifx\csname url\endcsname\relax
  \def\url#1{\texttt{#1}}\fi
\expandafter\ifx\csname urlprefix\endcsname\relax\def\urlprefix{URL }\fi
\providecommand{\eprint}[2][]{\url{#2}}

\bibitem[{Li}(2001)]{li:2001}
T.-P. {Li}, \emph{Chinese Journal of Astronomy and Astrophysics} \textbf{1},
  313--332 (2001), \eprint{arXiv:astro-ph/0109468}.

\bibitem[{Li} and {Muraki}(2002)]{li:2002}
T.-P. {Li}, and Y.~{Muraki}, \emph{Astrophysical Journal} \textbf{578},
  374--384 (2002), \eprint{arXiv:astro-ph/0204368}.

\bibitem[{Rizzuto} et~al.(2007)]{rizzuto:2007}
D.~{Rizzuto}, C.~{Guidorzi}, P.~{Romano}, S.~{Covino}, S.~{Campana},
  M.~{Capalbi}, G.~{Chincarini}, G.~{Cusumano}, D.~{Fugazza}, V.~{Mangano},
  A.~{Moretti}, M.~{Perri}, and G.~{Tagliaferri}, \emph{Monthly Notices of the
  Royal Astronomical Society} \textbf{379}, 619--628 (2007),
  \eprint{arXiv:0704.2486}.

\bibitem[{Shen} and {Song}(2003)]{shen:2003}
R.-F. {Shen}, and L.-M. {Song}, \emph{Publications of the Astronomical Society
  of Japan} \textbf{55}, 345--349 (2003), \eprint{arXiv:astro-ph/0301553}.

\end{thebibliography}

\IfFileExists{\jobname.bbl}{}
 {\typeout{}
  \typeout{******************************************}
  \typeout{** Please run "bibtex \jobname" to optain}
  \typeout{** the bibliography and then re-run LaTeX}
  \typeout{** twice to fix the references!}
  \typeout{******************************************}
  \typeout{}
 }

\end{document}